\documentclass[aps,prb,reprint,superscriptaddress]{revtex4-2}
\usepackage{hyperref}
\usepackage{graphicx}
\usepackage{color}
\usepackage{amsfonts}
\usepackage{amsmath}
\usepackage[utf8]{inputenc}
\usepackage[T1]{fontenc}
\usepackage{mathtools}
\usepackage{bbm}
\usepackage{amssymb}

\usepackage[dvipsnames]{xcolor}
\hypersetup{colorlinks=true, urlcolor=blue, citecolor = blue}

\begin{document}

\title{Signatures of Majorana bound states in scanning gate microscopy of hybrid nanowires}

\author{S. Maji}
\email{maji@agh.edu.pl}
\affiliation{AGH University of Krakow, Academic Centre for Materials and Nanotechnology, al. A. Mickiewicza 30, 30-059 Krakow, Poland}

\author{M. P. Nowak}        
\email{mpnowak@agh.edu.pl}
\affiliation{AGH University of Krakow, Academic Centre for Materials and Nanotechnology, al. A. Mickiewicza 30, 30-059 Krakow, Poland}

\date{\today}

\begin{abstract}
We theoretically study scanning gate microscopy of a superconductor-proximitized semiconducting wire focusing on the potential for detection of Majorana bound states. We exploit the possibility to create a local potential perturbation by the scanning gate tip which allows controllable modification of the spatial distribution of the Majorana modes, which is translated into changes in their energy structure. When the tip scans across the system, it effectively divides the wire into two parts with controllable lengths, in which two pairs of Majorana states are created when the system is in the topological regime. For strong values of the tip potential, the pairs are decoupled, and the presence of Majorana states can be detected via local tunneling spectroscopy that resolves the energy splittings resulting from the Majorana states wave functions overlap. Importantly, as the system is probed spatially via the tip, this technique can distinguish Majorana bound states from quasi-Majorana states localized on smooth potential barriers. We demonstrate that for weaker tip potentials, the two neighboring Majorana states hybridize, opening pronounced anticrossings in the energy spectra which are reflected in local conductance maps and which result in non-zero non-local conductance features. Finally, we demonstrate that the scanning gate microscopy technique can be used to discriminate between the trivial and topological nature of the zero-bias conductance peak in disordered wires.
\end{abstract}

\maketitle
\section{Introduction}
Scientific interest in the field of detecting Majorana bound states (MBSs) has grown abruptly in the last decade, as they are the key to generating topological qubits, which are the building blocks of fault-tolerant quantum computers \cite{Nayak,KITAEV20032,Alicea}. Particular effort has been put into the realization of MBSs in hybrid, quasi-one-dimensional semiconductor-superconductor systems \cite{Oreg,AYuKitaev,Sau,PhysRevB.82.214509,Lutchyn,PhysRevB.84.094505}, exploiting the mutual action of Zeeman interaction, Rashba spin-orbit coupling, and electron-electron pairing.

Spectroscopic conductance measurements, in which the conductance peak is typically sought for \cite{Mourik, Zhang2018}, are often employed to search for signatures of MBSs. However, it was realized that the sole observation of a peak at zero energy \cite{Gibertini} cannot be considered as an unambiguous hallmark of the presence of MBSs, since such peaks can also originate from the disorder present in the system \cite{PhysRevResearch.2.013377, PhysRevB.103.214502, Das} or from a specific potential profile in the vicinity of the electric gates that creates local excitations called quasi-Majorana states in an otherwise trivial system \cite{PhysRevB.86.100503, PhysRevB.97.214502, PhysRevB.98.155314, Christopher, Adriaan}.

The two-end MBSs wave functions decay exponentially into the wire region as $\Psi(x) \propto e^{-x/\xi}$, with the decay being determined by the coherence length $\xi$. Upon overlap of the MBSs wave functions localized on the opposing ends, the degeneracy of the energy levels of the states is lifted. Crucially, as the magnetic field is increased, the energy of overlapping MBSs starts to oscillate around zero with the amplitude increasing in the magnetic field. Possible observation of this feature was considered to be strong evidence of the presence of MBSs \cite{PhysRevB.86.220506}, but later it was realized that the orbital effects of the magnetic field diminish the oscillations \cite{PhysRevB.97.155409, PhysRevB.97.045419} and also that the MBSs are never observed in a sufficiently large span of the magnetic fields due to the closing of the parent superconducting gap \cite{Zhang2018}. On the other hand, the control of the oscillation amplitude via the global chemical potential is largely limited by a small regime of the chemical potential that guarantees the topological regime at small magnetic fields. Finally, the systematic control over the length of the wire is in practice only possible in atomic chain realizations \cite{Jäck2021}.

In this work, instead of varying the magnetic field, we explore an alternative route that can be used to exploit the oscillatory behavior of MBSs energy levels upon their overlap. We propose, locally perturbing the potential profile in the system using the scanning gate microscopy (SGM) technique---which allows tuning the spatial span of MBSs in a single experimental measurement. In SGM experiments, a charged atomic force microscope tip scans over the system inducing a local potential perturbation \cite{PhysRevB.77.125310, PhysRevLett.99.136807, Pala_2009, Aidala2007, Szafran}. As the SGM tip deflects the electron trajectories in open systems, this method is widely used to visualize electron flow and electron self-interference in semiconducting structures \cite{doi:10.1126/science.289.5488.2323, Topinka2001, PhysRevB.80.041303, Iagallo,  Aoki, LeRoy, 10.1063/1.1484548, PhysRevB.90.165303, PhysRevB.90.035301, PhysRevB.94.075301, PhysRevB.96.165310, Prokop_2020}. SGM is also used to affect the charging and electron localization in confined systems \cite{PhysRevB.79.134530, 10.1063/1.2787163, Michael, Denisov2022}. 
The SGM technique has been considered for superconducting systems both from a theoretical \cite{PhysRevB.106.035432, S.Maji} and experimental \cite{Bhandari2020} perspective with a very recent demonstration of SGM of supercurrent in Josephson junctions \cite{lombardi2025supercurrentmodulationinsbnanoflagbased}.

In our work, we theoretically study the application of the SGM technique to discriminate between the trivial and topological origin of the zero-bias peaks in a hybrid superconductor-semiconductor system. We show that the SGM applied to a quasi-one-dimensional proximitized wire can induce the localization of an additional Majorana pair \cite{PhysRevB.95.045429} and that the change of the position of the probe changes the overlap between MBSs localized on each side of the tip, which in turn induces conductance oscillations. We also show that the SGM technique, due to probing the system spatially, can distinguish MBSs from quasi-MBSs \cite{PhysRevB.86.100503, PhysRevB.97.214502, PhysRevB.98.155314, Christopher, Adriaan, PhysRevB.108.205426}, or disorder-induced zero-energy states, and that the variation of the tip potential allows hybridization of the MBSs from the pairs localized at the opposite side of the tip.

This paper is organized as follows. In Section II we introduce the theoretical model used to obtain the energy spectra and transport features of the system. In Sect. III A we demonstrate the SGM induced energy splitting of MBSs, Sect. III B discusses the quasi-MBSs case in the context of the SGM study, in Sect. III C we consider local and non-local conductance features of systems with weaker SGM potential tip and finally in Sect. III D we present results with disorder. Section IV provides a discussion of the potential realization of the proposed technique, while Sect. V summarizes our results.

\section{Theory}
\begin{figure}[h!]
    \centering
    \includegraphics[width=0.98\linewidth]{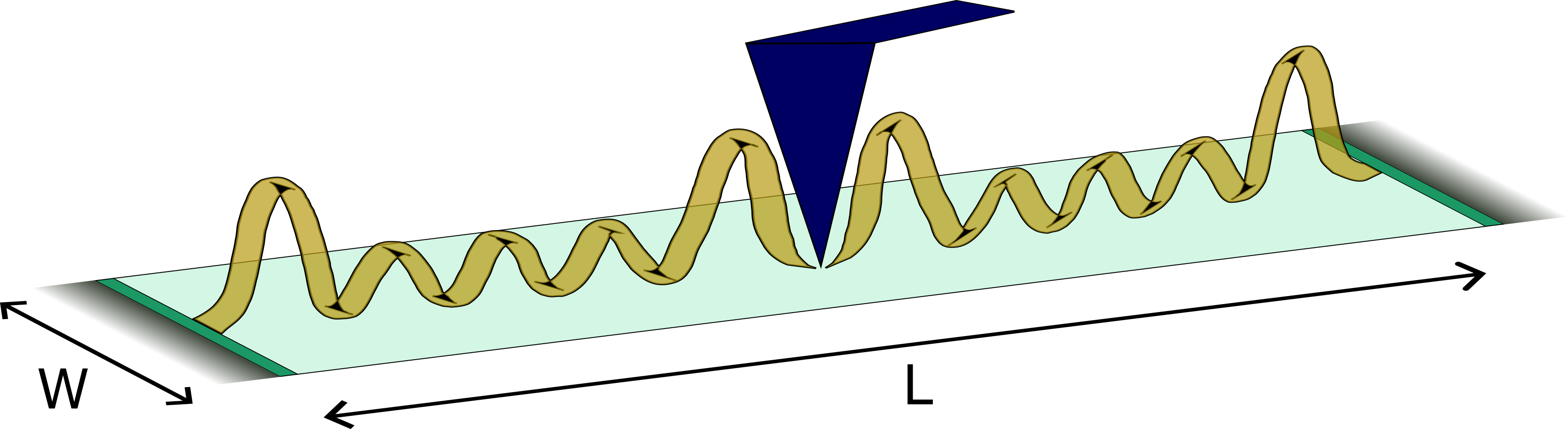}
    \caption{Scheme of the considered system. A proximitized stripe (green) that hosts MBSs (yellow) connected to two normal leads (gray) through tunneling barriers (dark green) is scanned by the atomic force microscopy tip (dark blue) which decouples the system into two parts with varied lengths.}
    \label{fig:system}
\end{figure}

We consider a model of a system of a quasi-one-dimensional proximitized semiconducting wire which is described by the Hamiltonian,
\begin{equation}
\begin{split}
\label{eqn:Hamiltonian}
H = &\left(\frac{\hbar^2\textbf{k}^2}{2m^*} + V_b(x,y) + V_{\mathrm{SGM}}(x,y) -\mu\right) \sigma_0\otimes\tau_z \\&+ \Delta\sigma_0\otimes\tau_x + \alpha(\sigma_x k_y - \sigma_y k_x)\otimes\tau_z + E_z\sigma_x\otimes\tau_0.
\end{split}
\end{equation}
$\Delta$ corresponds to the induced gap in the wire due to a nearby superconductor, $\alpha$ is the Rashba spin-orbit coupling amplitude, and $E_z=\frac{1}{2}g\mu_B B$, the Zeeman interaction term for the magnetic field oriented along the wire, $\sigma_i$ and $\tau_i$ with i = x, y, z are the Pauli matrices acting on
spin and electron-hole degree of freedom, respectively. Here, we use the model of a quasi-one-dimensional nanowire with integrated superconductivity to portray the proximitized semiconductor, but as we have checked, the same features are obtained for systems consisting of a bare semiconducting part and one or two proximitizing superconducting leads. 

The scheme of the system is presented in Fig. \ref{fig:system}. The scattering region in the stripe (light green) is connected to two normal leads (gray) through potential barriers $V_b$ (dark green in Fig. \ref{fig:system}) of width $20$ nm and height $V_g = 14.8$ meV (unless stated otherwise), which tune the electronic transport characteristics into the tunneling regime.
Finally, $V_{\mathrm{SGM}}$ is the SGM-induced potential in the system  \cite{Szafran},
\begin{equation}
V_{\mathrm{SGM}}(x,y)=\frac{V_{\mathrm{tip}}}{1+\frac{(x-x_{\mathrm{tip}})^2+(y-y_{\mathrm{tip}})^2}{d_{\mathrm{sgm}}^2}},
\end{equation}
where unless stated otherwise we set $V_\mathrm{tip}=50$ meV and $d_{\mathrm{sgm}}=50$ nm. The particular choice of SGM tip potential shape \cite{PhysRevLett.99.136807} or the specific model parameter values does not change the obtained results in a qualitative way \cite{Supplementary} with the exception of the $V_\mathrm{tip}$ value which can be used to couple neighboring MBSs as we show in the following. The spatial position of the SGM tip is determined by the pair of coordinates ($x_{\mathrm{tip}},y_{\mathrm{tip}}$). 

We adopt the system shape: length $L = 2000$ nm, width $W = 100$ nm, assume InSb material parameters $m^*=0.014m$, $\mu = 3$ meV, $g =-50$, $\alpha = 40$ meVnm, and the induced superconducting gap of $\Delta = 1$ meV corresponding to superconductors such as Nb, NbTiN \cite{Salimian, Zhi, PhysRevB.99.245302, Satchell}, however, the results obtained here are not specific to the choice of this particular value of parameters.

To infer the conductance spectroscopy features of the device, we calculate the zero-temperature conductance according to the formula
\begin{equation}
G_{ij}(E) = \frac{\partial I_i}{\partial V_j} = \frac{e^2}{h}(\delta_{ij}N_i^e(E)-T^{ee}_{ij}(E)+T^{he}_{ij}(E)).
\label{conductanceformula}
\end{equation}
$I_i$ is the current entering the scattering region from the terminal $i$ and $V_j$ is the voltage applied to the $j$'th lead. $N_i$ is the number of electronic modes in the $i$'th lead, $T^{ee}_{ij}(E)$ is the electron-to-electron transmission coefficient for electrons injected from the lead $j$ and captured at the terminal $i$, and $T^{he}_{ij}(E)$ is the corresponding electron-to-hole transmission coefficient \cite{Rosdahl}. Equation (\ref{conductanceformula}) allows the calculation of both: non-local ($i \ne j$) and local conductance ($i=j$), which we obtain by injecting electrons from the left lead and which we denote as $G \equiv G_{11}$ in the following.

The coefficients $T^{ee}_{ij}(E)$ and $T^{he}_{ij}(E)$ are obtained from the scattering matrix of the system calculated at the energy $E$ corresponding to voltage bias at the $j$'th lead,
\begin{equation}
    T^{\alpha,\beta}_{ij}(E) = \mathrm{Tr}\left( [S^{\alpha,\beta}_{ij}(E)]^\dagger S^{\alpha,\beta}_{ij}(E)\right).
\end{equation}
$S^{\alpha,\beta}_{ij}(E)$ is the scattering matrix block corresponding to the particles of type $\beta$ injected from the $j$'th lead and scattered back as particle type $\alpha$ into the $i$'th lead. 

The scattering matrix of the system is obtained by discretizing the Hamiltonian Eq. (\ref{eqn:Hamiltonian}) on a finite mesh with lattice spacing $a = 10$ nm and solving the transport problem using the Kwant package \cite{ChristophKwant}. Conductance maps are calculated with the help of the Adaptive Python package \cite{nijholt_2023_10215599}. The code used to obtain the results presented in this paper is available in an online repository \cite{maji_2025_17264576}.

\section{Results}
\subsection{Energy splitting as a signature of MBSs presence}
\begin{figure}[h!]
    \centering
    \includegraphics[width=0.68\linewidth]{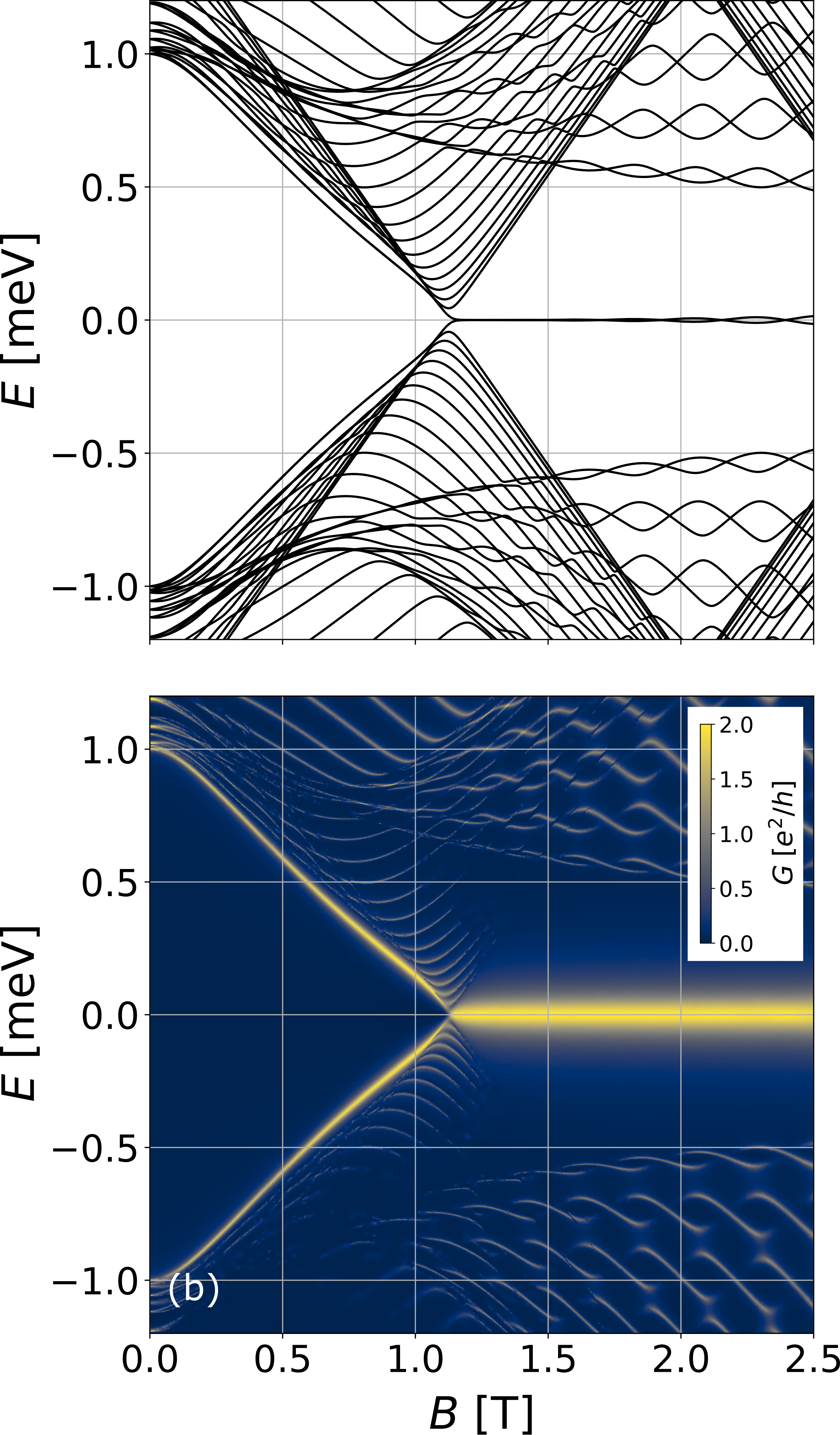}
    \caption{(a) Energy spectrum versus the magnetic field of a proximitized wire and the respective local conductance map (b) obtained without the SGM tip.}
    \label{fig:noSGM}
\end{figure}

In Fig. \ref{fig:noSGM}(a) we show the energy spectra of a finite wire (without the leads) versus the magnetic field. A clear closing and reopening of the superconducting gap is observed at around $B = 1.1$~T, which is accompanied by the appearance of zero-energy Majorana modes after the reopening.

The spatial span of MBSs is determined by their coherence length as derived for a quasi-one dimensional nanowire \cite{PhysRevB.86.220506} in the presence of the magnetic field and spin-orbit coupling,
\begin{equation}
\xi \approx \frac{1}{\alpha\Delta}\sqrt{\left(\frac{\hbar^2}{m^*}\mu+\alpha^2\right)^2+\left(\frac{\hbar^2}{m^*}\right)^2(E_z^2-\Delta^2-\mu^2)}.
\label{eq:xi}
\end{equation}

For the parameters taken here and with $B=2$~T, the coherence length is estimated as $\xi \simeq 413$~nm. The MBSs are mostly localized close to the barriers. As $L \gg \xi$ the edge state wave functions are well separated and their lack of overlap is accompanied by almost complete degeneracy at zero energy of their energy levels seen in the spectra of Fig. \ref{fig:noSGM}. Consequently, in tunneling spectroscopy of Fig. \ref{fig:noSGM}(b) we observe a stable $2e^2/h$ peak at zero energy after the gap closing and reopening.

\begin{figure}[h!]
    \centering
    \includegraphics[width=0.98\linewidth]{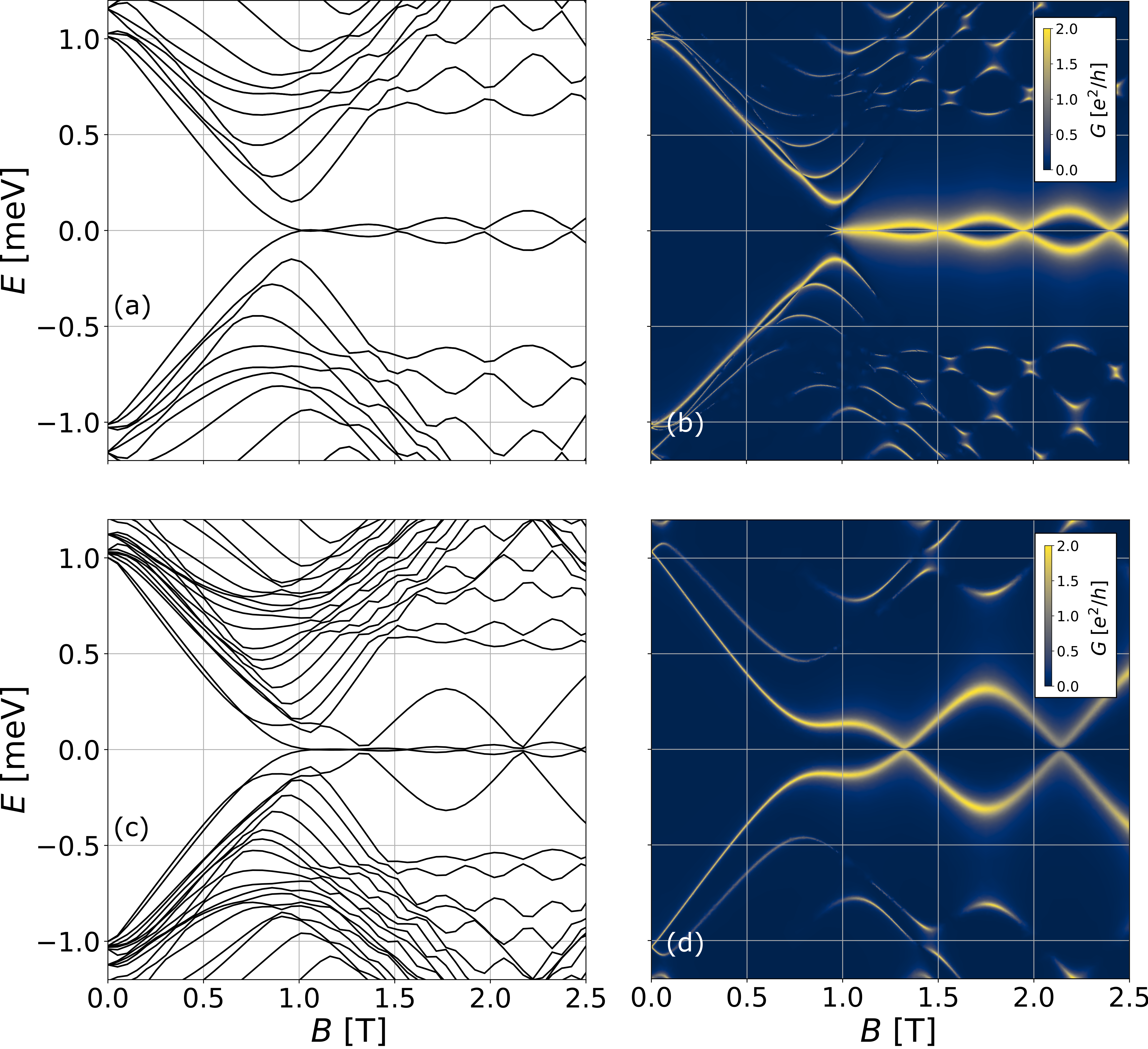}
    \caption{(a) and (c) the energy spectra of a proximitized wire versus the magnetic field in the presence of the SGM tip. (b) and (d) the local conductance maps in the tunneling regime. The top row is obtained for the SGM tip located at $x_{\mathrm{tip}} = 1000$~nm while the bottom on for $x_{\mathrm{tip}} = 500$ nm.}
    \label{fig:NSvsB}
\end{figure}

Let us now consider the case where the SGM tip is introduced into the system. In Fig. \ref{fig:NSvsB}(a), we show the energy spectrum with the tip located in the middle of the wire with $x_{\mathrm{tip}} = 1000$~nm and $y_{\mathrm{tip}} = 50$~nm.  Compared to Fig. \ref{fig:noSGM}(a), we observe a clear splitting of the energy levels corresponding to the MBSs with the magnitude of the splitting increasing as the magnetic field becomes stronger. The splitting is also clearly seen on the conductance map in Fig. \ref{fig:NSvsB}(b). Note that each of the energy levels oscillating around zero energy is actually degenerate and corresponds to two pairs of MBSs located on each side of the wire divided by the potential tip [see the schematic depiction in Fig. \ref{fig:system}]. This becomes apparent when we move the tip to $x_{\mathrm{tip}} = 500$ nm. The corresponding energy structure is presented in Fig. \ref{fig:NSvsB}(c). We clearly observe two sets of curves oscillating around zero energy. They correspond to two pairs of MBSs created on each side of the wire which are separated by the repelling potential of the tip. The pair of MBSs located in the longer segment, $x > 500$ nm, has a smaller overlap, hence the oscillations have a smaller amplitude, whereas the states in the pair located in the shorter segment (with length comparable to $\xi$) strongly overlap giving oscillations with an almost three-fold longer period and a significant amplitude. 

The local conductance calculated from the left lead is shown in Figs. \ref{fig:NSvsB}(b) and (d). For the case of $x_{\mathrm{tip}} = 1000$~nm, the tunneling spectroscopy directly reproduces the energy spectrum of panel (a) with oscillating energy levels close to zero energy after the topological transition. However, the situation is different for $x_{\mathrm{tip}} = 500$ nm. From the two oscillating lobes around zero energy seen in the spectrum of Fig. \ref{fig:NSvsB}(c) the spectroscopy map Fig. \ref{fig:NSvsB}(d) reveals only one---that corresponds to the MBSs localized on the shorter side of the wire, where the conductance spectroscopy is obtained. Note that in this case, the non-local spectroscopy is zero through the whole $E, B$ plane as the tip blocks the transport along the wire.

\begin{figure}[h!]
    \centering
    \includegraphics[width=0.98\linewidth]{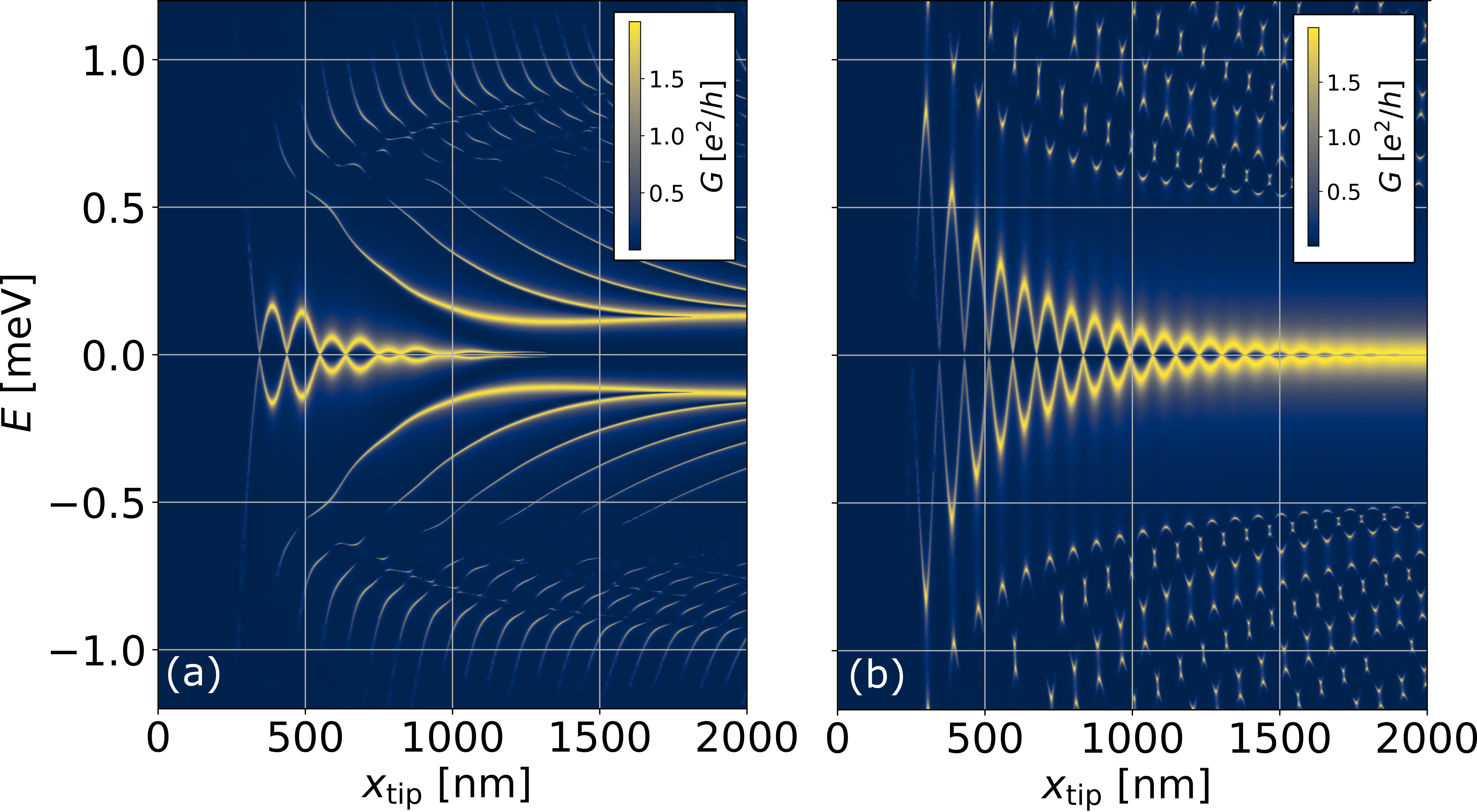}
    \caption{Conductance spectroscopy maps versus the position of the tip along the wire for (a) $B$ = 1 T and (b) $B$ = 2 T.}
    \label{fig:NSvsxtip}
\end{figure}

\begin{figure*}[ht!]
    \centering
    \includegraphics[width=1\linewidth]{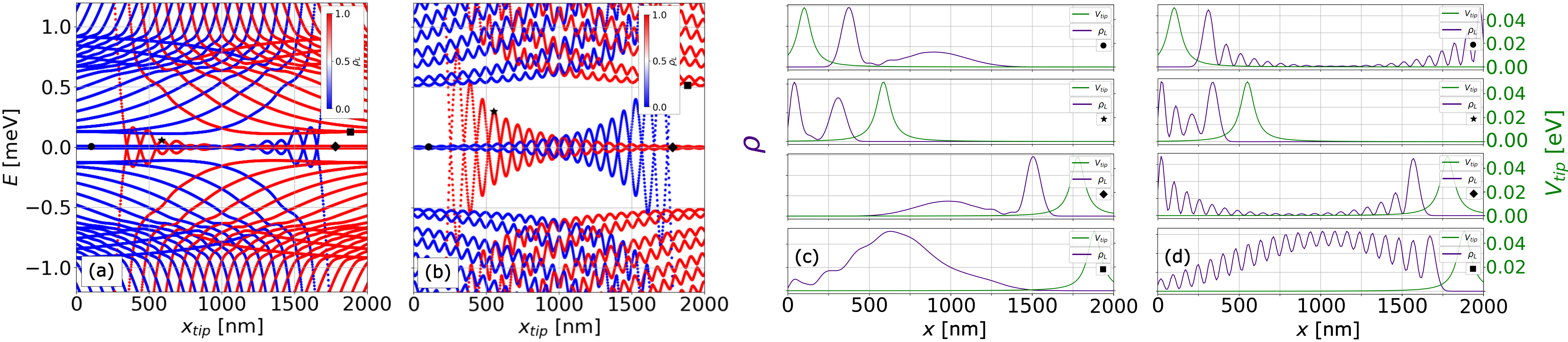}
    \caption{(a) and (b) the energy spectra versus the position of the tip along the wire. The colors denote the weight of the probability density located on the left side of the tip. (c) and (d) with violet show the cross-section of the probability density for $y = 50$~nm for the states whose corresponding energy levels are denoted with symbols in panels (a) and (b). The green curves in (c) and (d) depict the potential profile in the wire due to SGM tip. Panels (a) and (c) correspond to $B = 1$~T while (b) and (d) to $B = 2$~T.}
    \label{fig:DOSlocalization}
\end{figure*}

Let us now consider a case where the magnetic field is set constant, but the tip scans across the wire length. Figure \ref{fig:NSvsxtip} shows the conductance spectroscopy maps of the wire versus the position of the tip at two values of the magnetic field. In Fig. \ref{fig:NSvsxtip}(b), which is obtained for $B = 2$ T, we observe that as the tip moves from the left edge of the wire along the $x$ axis, the conductance oscillates with a constant period and with a pronounced decrease in the oscillation amplitude. This is in line with the expected dependence of the MBS energy splitting on the length of the system with the energy splitting $\Delta E \propto \hbar^2k_F\exp{[-2L/\xi]}/m^*\xi\cdot\cos(k_FL)$, where $k_F$ is the effective Fermi wave vector associated with the MBSs \cite{PhysRevB.86.220506}. This can be further confirmed by inspecting the energy spectrum and the distribution of the wave function in the system. Figures \ref{fig:DOSlocalization}(a) and (b) show the energy spectrum of the wire without the leads, with the colors denoting the weight of the probability density located on the left side of the tip,
\begin{equation}
\rho_L = \int_{-W/2}^{W/2}\int_0^{x_{tip}} \sum_{q = e/h,\sigma = \uparrow/\downarrow} |\Psi_{q,\sigma}(x,y)|^2dxdy.
\end{equation}

In Fig. \ref{fig:DOSlocalization}(b) we observe that the spectrum is symmetric with respect to $x = 1000$ nm, but the conductance spectroscopy [Fig. \ref{fig:NSvsxtip}(b)] only resolves the states located on the left side of the tip---with the energy levels colored red. This particular structure of the energy spectrum can be understood by inspecting the probability densities of the corresponding states.

Figure \ref{fig:DOSlocalization}(d) shows in violet the probability density of the states whose energy levels are denoted in panel (b) by the corresponding symbols. The state whose energy level is denoted by $\bigstar$ corresponds to the MBSs [violet curve in Fig. \ref{fig:DOSlocalization}(d)] located between the tunneling barrier and the SGM tip potential [green curve in \ref{fig:DOSlocalization}(d)]. The high overlap of the MBSs leads to a significant deviation of their energy levels from zero energy. As $x_{\mathrm{tip}}$ increases, the amplitude of the oscillations of the energy levels of the states located on the left side of the tip diminishes, resulting from the pronounced delocalization of the two MBSs [see the probability density for the state denoted with $\blacklozenge$ in Fig. \ref{fig:DOSlocalization}(d)]. The similar situation occurs for the states located on the right side of the tip, denoted by blue colors in Fig.~\ref{fig:DOSlocalization}(b) with the density shown in Fig.~\ref{fig:DOSlocalization}(d) with the $\bullet$ symbol. Finally, the ribbon-shaped states with energy levels higher than those corresponding to MBSs are ordinary Andreev-bound states localized in the bulk of the junction [see the probability density of the $\blacksquare$ state in Fig. \ref{fig:DOSlocalization}(d)].

Let us now switch to the case of a smaller magnetic field $B = 1$~T. If we look at the energy spectrum of Fig.~\ref{fig:noSGM}(a) we see that at $B = 1$~T the gap is not yet closed and the system {\it without} the SGM tip is in the trivial regime. Despite that, in the conductance map of Fig. \ref{fig:NSvsxtip}(a), we do observe resonances on states oscillating around zero energy. They are also visible in the energy spectra in Fig. \ref{fig:DOSlocalization}(a). 

For the magnetic field value below the topological transition, when the condition (written for a one-dimensional system) $E_z > \sqrt{\Delta^2 + \mu^2}$ is not yet fulfilled, the bulk of the system remains trivial, but the smooth tail of the SGM tip potential [see the green curves in Fig. \ref{fig:DOSlocalization}(c)] induces the creation of quasi-MBS \cite{PhysRevB.86.100503, Adriaan, PhysRevB.97.214502, PhysRevB.98.155314, Christopher} that in this case are strongly localized near the tip. This is exactly what we observe in the spectra of Fig. \ref{fig:DOSlocalization}(a) and the density of states denoted by $\bullet$, $\bigstar$ and $\blacklozenge$. Quasi-MBSs have zero energy when the tip is located at the left edge of the system [see the density of $\bullet$ state in Fig. \ref{fig:DOSlocalization}(c)], while their energy levels slightly split from zero when they are located in a small region between the tunneling barrier and the tip potential [$\bigstar$ in Fig. \ref{fig:DOSlocalization}(c)]. Importantly, as the tip moves to the right, the quasi-Majorana follows its localization, becoming decoupled from the left edge of the system [see the plot of the density denoted by $\blacklozenge$ in Fig. \ref{fig:DOSlocalization}(c)]. This is in contrast to the situation when the bulk of the system was in the topological regime [see $\blacklozenge$ state of Fig. \ref{fig:DOSlocalization}(d)] and despite a similar structure of the energy levels of quasi-MBSs and true-MBSs, the conductance spectra probed with the SGM bear significant differences. For the case of quasi-MBS, the resonance close to zero energy vanishes for the tip located on the right side of the system [Fig. \ref{fig:NSvsxtip}(a)], while they are present for the fully topological system [Fig. \ref{fig:NSvsxtip}(b)] allowing distinguishing those two scenarios. Furthermore, the distinction between the two cases is clear when one compares the spectra obtained without the SGM tip in Fig. \ref{fig:noSGM}(b) where for $B = 1$ T we observe still open trivial gap, while for $B = 2$ T the gap is past the topological transition.

\subsection{Smooth potential barrier---quasi-MBSs case}
\begin{figure}[h]
    \centering
    \includegraphics[width=0.68\linewidth]{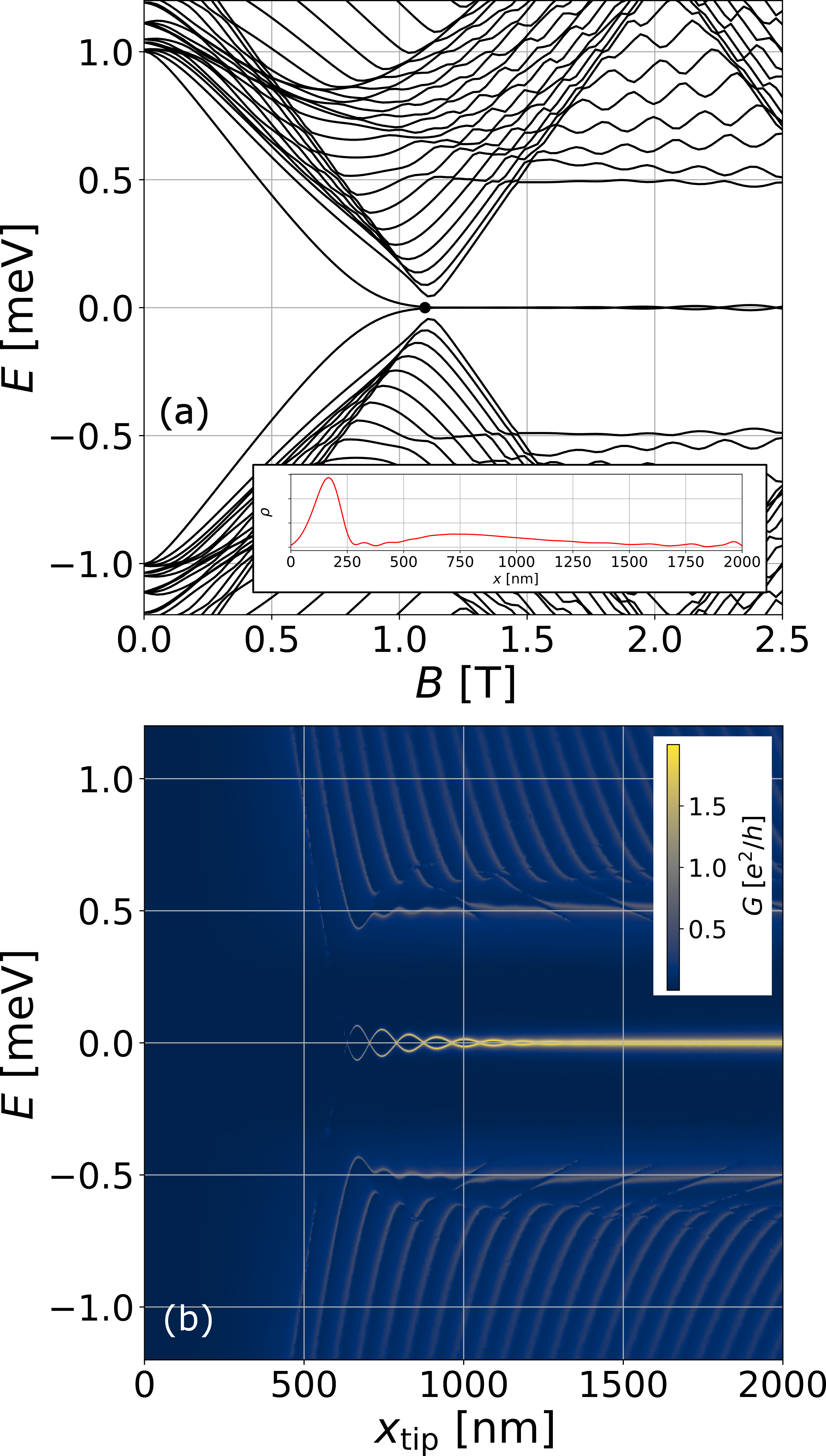}
    \caption{(a) Energy spectrum versus the magnetic field for the wire with smooth potential barrier and without the SGM tip. (b) Conductance map for $B = 1.1$ T versus the tip position.}
    \label{fig:quasimajorana}
\end{figure}

\begin{figure*}[ht!]
    \centering
    \includegraphics[width=0.7\linewidth]{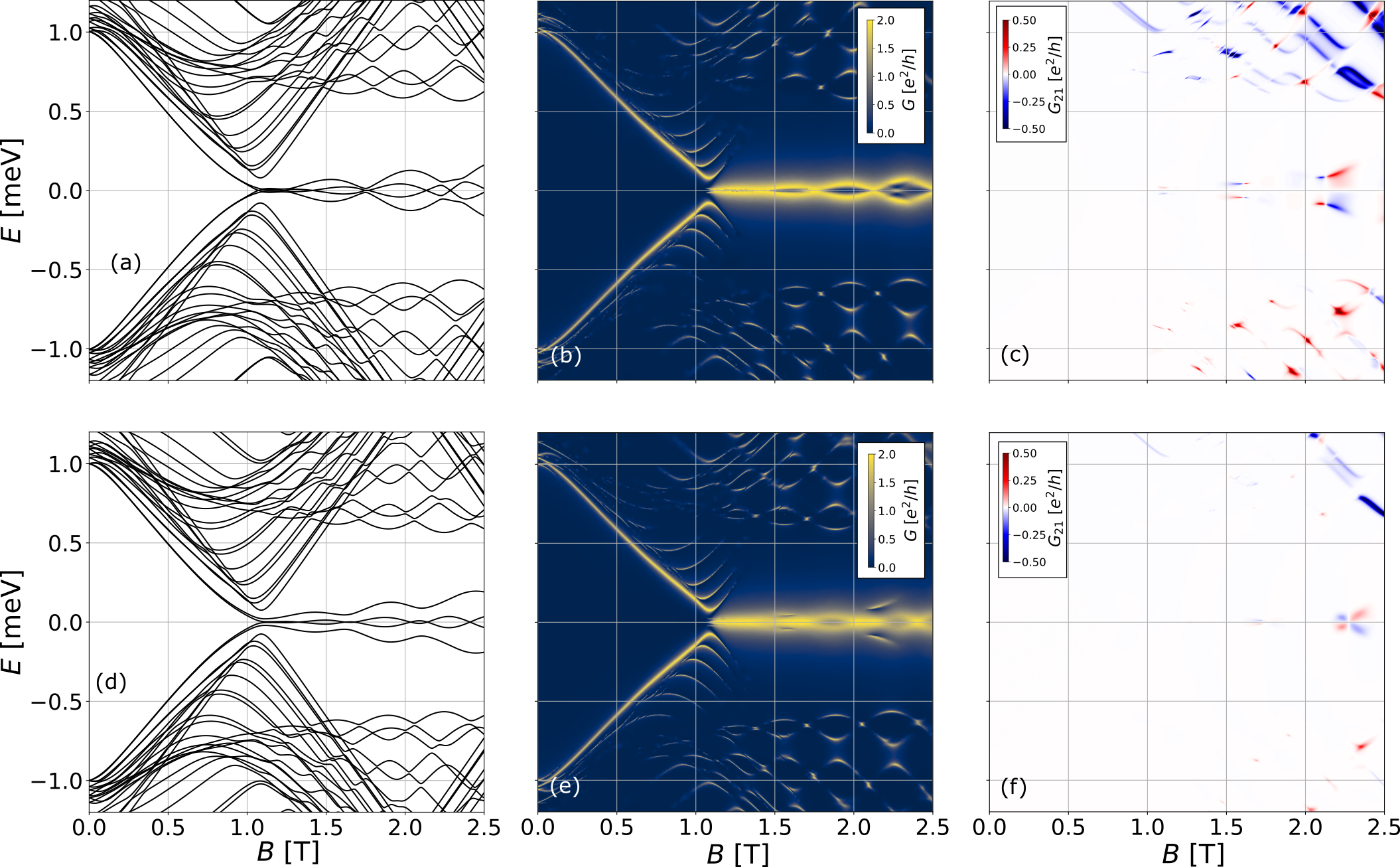}
    \caption{(a) and (d) energy spectra versus the magnetic field for a system with lower tip potential values. The corresponding local (b), (e) and non-local (c), (f) conductance maps. (a)-(c) are obtained for $V_\mathrm{tip}= 10$ meV while (d)-(f) for $V_\mathrm{tip} = 7.5$ meV. $x_{\mathrm{tip}} = 1200$ nm for all the plots.}
    \label{fig:smallervtip}
\end{figure*}

The observed quasi-MBSs can already appear in systems without the SGM tip and constitute a serious obstacle to distinguish the trivial or topological superconductivity nature of the bulk of the system \cite{PhysRevB.86.100503, PhysRevB.97.214502, PhysRevB.98.155314,  Adriaan, Christopher}. They can appear in systems where the barriers at the ends of the wire are smooth enough, which can often happen due to dielectric layers separating the nanowire and underneath metallic gates.

To inspect such a case let us replace the sharp tunneling barrier on the left side of the system considered so far by a smooth one described by the equation
\begin{equation}
V_{b}(x)=V\exp[-\frac{(x-x_0)^2}{2\sigma^2}].
\end{equation}
The Gaussian-shaped barrier has the height of $V = 3.01$ meV, $x_0 = 0$ is its center, and the smoothness is determined by $\sigma$ = 200 nm. The energy spectrum versus the magnetic field for the system with the smooth barrier is shown in Fig.~\ref{fig:quasimajorana}(a). We observe that two Andreev-bound states coalesce as the magnetic field increases and merge even before the gap is closed by the rest of the Andreev-bound states. In the inset of Fig. \ref{fig:quasimajorana}(a) we show the probability of the zero-energy state denoted by the dot in the energy spectrum, just before the bulk gap closes and reopens. We observe the localization of the quasi-MBS in the vicinity of the potential barrier at the left edge of the system. In Fig. \ref{fig:quasimajorana}(b) we show the conductance map versus the position of the SGM tip. When the tip is close to the left edge, it blocks transport, and the conductance is zero. Upon increasing $x_\mathrm{tip}$, we observe an oscillatory splitting of the energy levels close to zero. The splitting, however, quickly vanishes as the tip moves to the right as a result of the strong localization of the quasi-MBSs on the left-hand side of the system. This is in contrast to the case of true-MBSs presented in Fig. \ref{fig:NSvsxtip}(b) where not only the oscillations had a significantly larger amplitude due to the localization of two edge modes between the tunneling barrier and the tip potential (resulting in their pronounced overlap---see the probability density of the state denoted with the $\bigstar$ symbol in Fig. \ref{fig:DOSlocalization}(d)) but also endure to the large values of $x_\mathrm{tip}$ reaching the whole length of the wire. 

Note that conductance spectroscopy does not resolve states in the window between the zero-energy mode and $|E| < 0.5$ meV, despite their presence at $B = 1.1$ T seen in the spectrum of Fig. \ref{fig:quasimajorana}. We have checked that the wave function of those Andreev bound states is strongly suppressed at the location of the potential barrier, virtually making them decoupled from the left lead, suppressing the electron tunneling to them.

\subsection{Mixing of nearby MBSs}
The large value of $V_\mathrm{tip} = 50$ meV results in the creation of a pronounced positive potential island in the wire, which virtually splits the system into two halves and results in the creation of two decoupled pairs of MBSs. The tip potential characterized here by the parameter $V_\mathrm{tip}$ is, in the experimental scenario, controlled by the charge accumulated on the tip \cite{Szafran} and can be controlled by the voltage applied to the tip \cite{PhysRevB.62.5174}.

In Figs. \ref{fig:smallervtip}(a) and (d) we show the energy spectrum for a system with a much smaller SGM potential value, namely $V_\mathrm{tip} = 10$ meV for (a) and $V_\mathrm{tip} = 7.5$ meV for (d). The tip is located at $x_{\mathrm{tip}} = 1200$ nm. We observe that as the potential induced by the SGM tip decreases, the crossings between the energy levels of the two pairs of MBSs are replaced by energy-level repulsions. This is particularly visible in Fig. \ref{fig:smallervtip}(d) at larger values of the magnetic field. There are pronounced anticrossings between the energy levels corresponding to the states located on the left side of the tip (with smaller oscillation amplitude) and those located on the right side. The energy level repulsion results from the mixing of the MBSs wave functions located on both sides of the tip. The corresponding local conductance maps are shown in panels (b) and (e) of Fig.~\ref{fig:smallervtip}. We observe that this time conductance spectroscopy reveals not only the energy levels of the MBSs localized on the side from which the transport is measured but also it possesses signatures of the energy levels of the MBSs localized on the other side of the tip---whenever the energy levels of both MBSs are close in energy---see the features appearing close to zero energy at $B > 2$~T in both maps.

\begin{figure}
    \centering
    \includegraphics[width=0.98\linewidth]{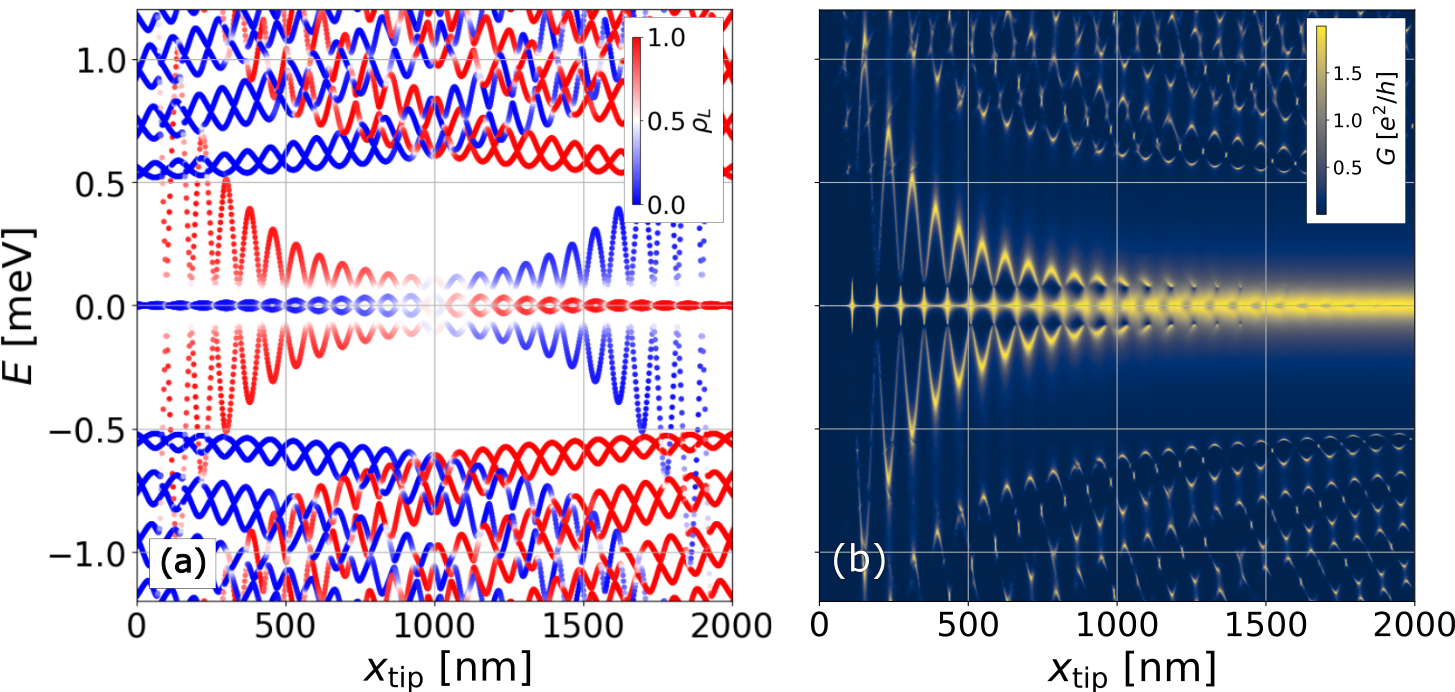}
    \caption{(a) The energy spectra versus the position of the tip along the wire. The colors denote the weight of the probability density located on the left side of the tip. (b) Tunneling spectroscopy measured from the left lead. The results are obtained for $V_\mathrm{tip} = 7.5$ meV.}
    \label{fig:dosmixing}
\end{figure}

Figure \ref{fig:dosmixing}(a) shows the energy spectrum versus the position of the tip for a system with weak tip potential $V_\mathrm{tip} = 7.5$ meV. Despite the similarities with the spectrum obtained for the $V_\mathrm{tip} = 50$ meV case [Fig. \ref{fig:DOSlocalization}(b)] we now observe that close to zero energy, the energy levels of MBSs localized on the left side of the tip (red) mix with those localized on the right side (blue). The mixing and the lack of definite localization at the left/right side of the tip are especially visible for the tip located in the vicinity of the middle of the system, where we observe faint colors (approaching white) of the lines corresponding to delocalization of the states among the whole length of the structure. This delocalization is translated into the spectroscopic features presented in Fig. \ref{fig:dosmixing}(b). The conductance measured from the left side of the system not only resolves the features of the MBS localized on the left side of the tip but is also sensitive to the other branch, as seen by the oscillating features close to zero energy for small values of $x_\mathrm{tip}$.

Finally, Figs. \ref{fig:smallervtip}(c) and (f) show the non-local conductance spectroscopy $G_{21}$. We observe that at a large magnetic field, where the overlap of MBSs is strong, the map shows non-zero values of the non-local signal. For larger values of the tip potential, the non-local spectroscopy is always zero because the tip potential made an impenetrable barrier for the charge carriers. Here, the situation is different; not only does the smaller value of the tip potential allow electron tunneling through the whole length of the wire but also the resonances close to zero energy signify that the tunneling transport occurs through two pairs of mutually overlapping MBSs \cite{PhysRevB.108.205426} resulting in a mostly electron (hole) transport for the negative (positive) non-local conductance values according to Eq.~(\ref{conductanceformula}).

\subsection{Disordered wire}

In realistic devices, one of the key limiting factors in the realization of the topological phase is disorder, which significantly limits the mean free path of devices to the order of $l_e \simeq 200-500$ nm for nanowires \cite{Plissard2012, Mourik, Shen} or $l_e \simeq 600$ nm for planar systems \cite{Fornieri,Banerjee,PhysRevB.107.245304}.

To assess the impact of the disorder on the possibility of SGM probing of MBSs, we now consider a case with a wire with a limited mean free path. The finite mean free path in the system is implemented through the inclusion of an onsite potential uniformly distributed within the range $[-U_d/2, U_d/2]$ \cite{Ando}, with the amplitude
\begin{equation}
    U_d=\mu \sqrt{\frac{6\lambda_F^3}{\pi^3a^2l_e}}.
\end{equation}
Here, $a$, $\lambda_F$, and $l_e$ are the lattice constant, the Fermi wavelength, and the mean free path, respectively, with $\lambda_F=2\pi\hbar/\sqrt{2m^*\mu}$.

\begin{figure}[h!]
    \centering
    \includegraphics[width=0.98\linewidth]{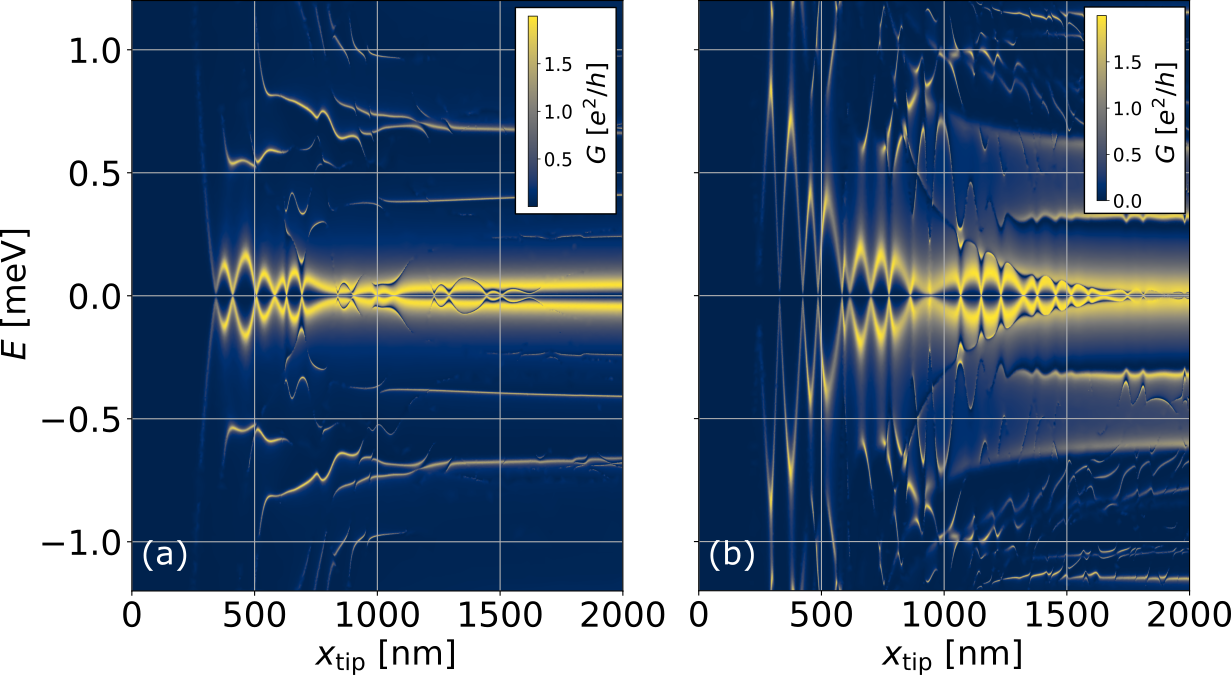}
    \caption{Conductance spectroscopy versus the tip position for $B = 1$ T (a) and $B = 2$ T (b) of a disordered system with $l_e = 200$ nm.}
    \label{fig:disorder}
\end{figure}

Figure \ref{fig:disorder} shows the conductance spectroscopy maps (obtained with sharp tunneling barriers on both sides of the wire) versus the tip position for $B = 1$ T (a) and $B = 2$ T (b) obtained for the disordered system with the mean free path $l_e = 200$ nm. For the case of $B = 1$ T where we deal with an initially trivial system, we observe a small disturbance of the oscillations as compared to the disorder-free case of Fig. \ref{fig:NSvsxtip}(a) as the quasi-MBSs are localized in a small region next to the SGM tip with a wave function span much smaller than the mean free path. When the magnetic field is increased to 2 T and the tip moves from the left edge of the wire, we observe in Fig. \ref{fig:disorder}(b) large-amplitude oscillations similarly to Fig. \ref{fig:NSvsxtip}(b) which is due to localization of the MBSs in a short region between the tunneling barrier and the tip potential (see the MBSs density of states plotted in Fig. \ref{fig:DOSlocalization}(d) with the $\bigstar$ symbol). Only when the tip is moved further to the right, we observe a strong disturbance of the oscillations as the MBS wave function spreads at the length $\xi > l_e$ and is strongly perturbed by the disorder along the wire.\\

\begin{figure}[h!]
    \centering
    \includegraphics[width=0.9\linewidth]{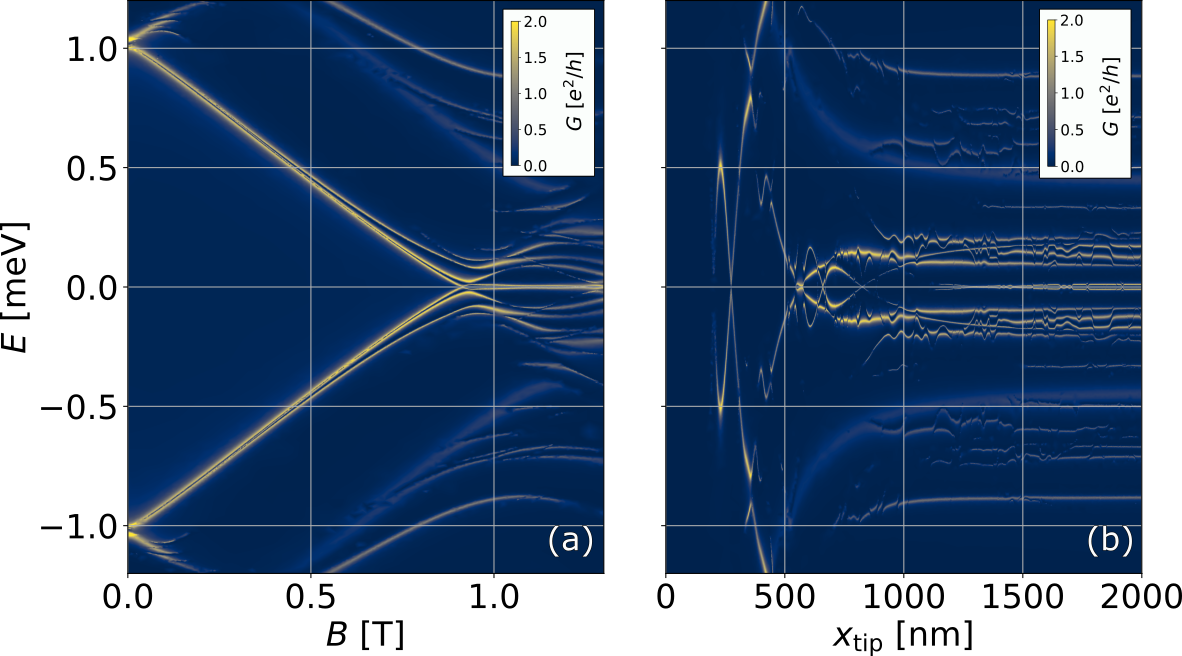}
    \caption{Conductance map without the SGM tip versus the magnetic field for uniformly disordered wire with the mean free path $l_e = 50$ nm with the disorder realization selected give a trivial zero bias conductance peak. (b) Conductance map for a constant magnetic field $B$ = 1.1 T, varying the tip position.}
    \label{fig:uniform_disorder}
\end{figure}

In the presence of considerable disorder, Andreev Bound State can coalesce into a zero energy state of trivial origin, which can result in a false-positive sign of MBS \cite{Das}. The wavefunction of those zero energy states, however, does not possess the crucial property of MBS: the localization at the wire ends with exponential decay into the bulk of the wire. Therefore, the proposed techniques can be used to discriminate between the topological and trivial origins of the zero bias conductance peak.

\begin{figure*}[ht!]
    \centering
    \includegraphics[width=1.0\linewidth]{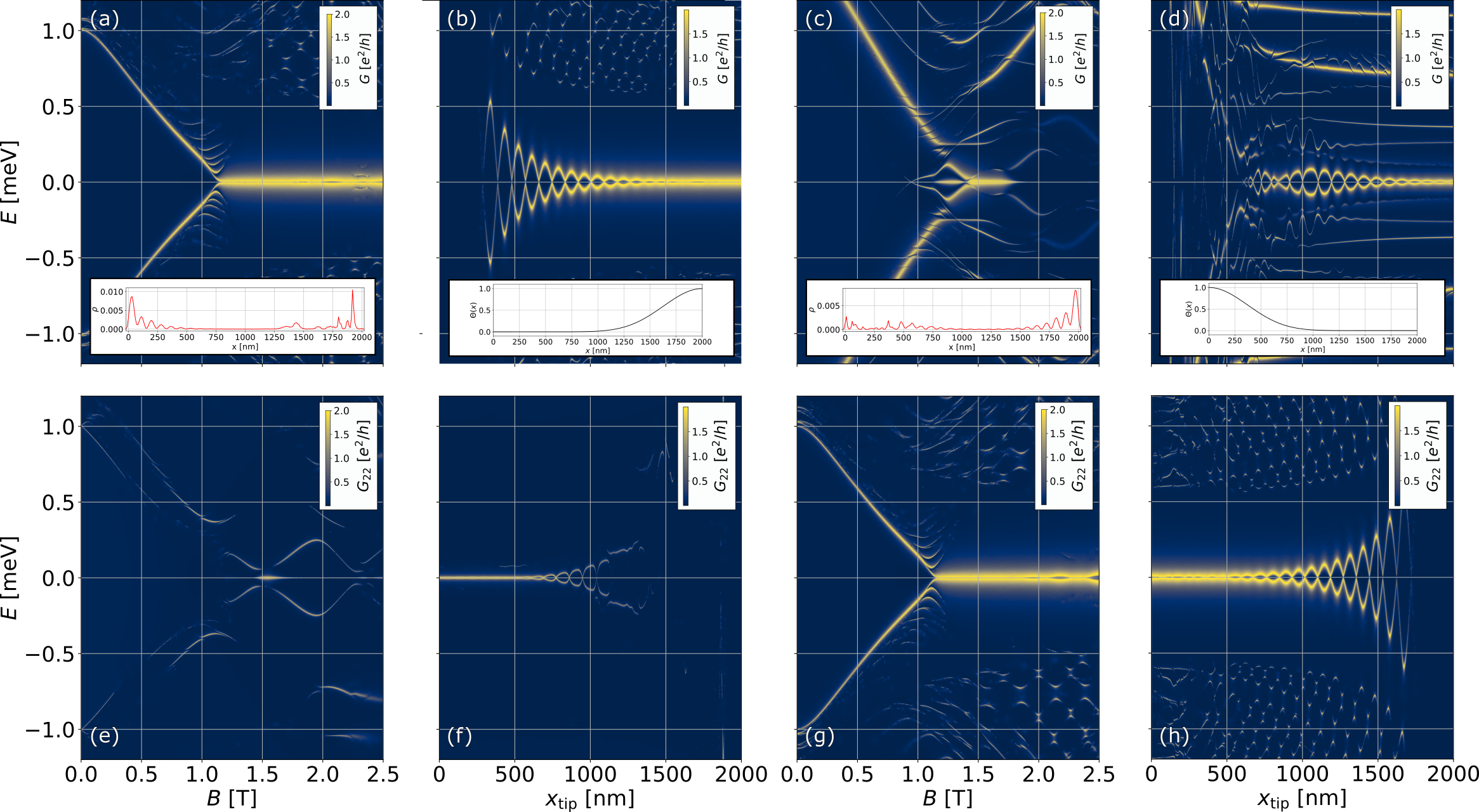}
    \caption{Conductance plots for spatially varying disorder with the amplitude modulated by function $\Theta$ displayed in the insets to panels (b) and (d). (a) and (c) conductance versus the magnetic field without the SGM tip. (b) and (d) conductance map for a constant magnetic field $B = 1.6$ T versus the tip position along the wire. For top panels (a-d) the conductance is obtained by tunneling from the left side of the system, while (e-h) correspond to the conductance obtained by tunneling through the right barrier.}
    \label{fig:Conductance_modulated_disorder_SGM_both_side}
\end{figure*}

We calculate the conductance spectroscopy of a strongly disordered wire, selecting a disorder realization that provides a trivial zero bias conductance peak---see Fig. \ref{fig:uniform_disorder}(a). The trivial nature of the system can be proven by the calculation of the topological charge \cite{PhysRevLett.106.057001} or topological visibility \cite{PhysRevB.94.035143}  $\mathcal{Q} = \mathrm{det}(r)$ (where $r$ is the reflection part of the scattering matrix), which has been used to quantify MBSs in disordered systems \cite{PhysRevB.108.085416}. For the considered system $Q = \mathrm{sign}(\mathcal{Q})$ at zero energy, yields the topological invariant \cite{PhysRevB.83.155429} and $Q = 1$ within the range of the considered magnetic field here (despite the presence of a zero-bias conductance peak). The trivial origin of the zero bias conductance peak in the highly disordered case, can be experimentally demonstrated by the SGM technique. In Fig. \ref{fig:uniform_disorder}(b) we show the conductance map versus the tip position along the wire obtained for $B = 1.1$ T. The absence of clear oscillations confirms that the zero bias conductance peak does not originate from the presence of MBSs compatible with the positive topological invariant. The evolution of the energy spectrum with varied disorder strength along with topological visibility ($\mathcal{Q}$) and with corresponding SGM conductance maps are further shown in the Supplementary section \cite{Supplementary}.\\

We also consider nanowires with nonuniformly distributed disorder (which could arise from, e.g., a specific shape of the gates that are used to tune the chemical potential of the system). We implement a disorder in the system that is modulated by a function $\Theta(x)$, which leaves one part of the system clean (where the topological phase can arise) and another part disordered. The disorder is introduced on the right side of the nanowire by the function $\Theta(x) = \sin(2\pi x/4L)^{12}$ and by $\Theta(x) = \cos(2\pi x/4L)^{12}$ for the left side of the wire (see the insets of Fig. \ref{fig:Conductance_modulated_disorder_SGM_both_side} (b) and (d)) and we take $l_e = 20$ nm.

Figure \ref{fig:Conductance_modulated_disorder_SGM_both_side}(a) shows the conductance versus the magnetic field for the disordered segment located on the right-hand side. As the left side of the system remains clean, MBS develops on the left side of the system (see the inset), and the conductance spectroscopy obtained for tunneling from the left contact sees a well-developed zero-bias conductance peak, which is associated with pronounced conductance oscillations when the system is scanned by the tip. Crucially, the oscillations are not affected by the presence of the disorder, as the SGM affects only the topological part in the clean portion of the wire. The situation is very different in the case of the disordered segment located on the left side of the nanowire, from where the tunneling spectroscopy is performed. Here the MBS on the left-hand side couples to the states localized on the disordered segment, and hence the conductance spectroscopy shows a rich structure of zero-bias conductance peak coupling to disorder bound states---Fig. \ref{fig:Conductance_modulated_disorder_SGM_both_side}(c). Most importantly, when we introduce the SGM tip and scan along the wire, we see a lack of conductance oscillation when the tip moves over the part of the wire where the disorder is present. This is due to the inability to develop a topological segment on the left-hand side of the wire---between the tunneling barrier and the SGM tip due to disorder. Only when the tip moves into the clean regime of the wire we observe the conductance oscillations [Fig. \ref{fig:Conductance_modulated_disorder_SGM_both_side} (d)]. 
The bottom panels of Fig. \ref{fig:Conductance_modulated_disorder_SGM_both_side} show the corresponding conductance obtained by tunneling not from the left side of the system but from the right one. The measurement from the other side of the wire, unlike Fig. \ref{fig:Conductance_modulated_disorder_SGM_both_side}(a), lacks a clear signature of the presence of a zero energy state---see Fig. \ref{fig:Conductance_modulated_disorder_SGM_both_side}(e). This could be taken as a false negative signature of a lack of a topological state in the system. However, this is not true, and we propose to use the SGM technique, which gives access to probing the interior of the structure to reveal the true nature of the state in the wire. Fig. \ref{fig:Conductance_modulated_disorder_SGM_both_side}(b) shows that there are clear oscillations of the conductance present when the tip scans the sample, consistent with the presence of MBS in the system. Similarly, for the disorder present on the other part of the wire visualized in Fig. \ref{fig:Conductance_modulated_disorder_SGM_both_side}(c-d), (g-h), we also find that the correlated measurements from both sides give false negative results for the presence of MBS, while the system is, in fact, in the topological regime with clear oscillations of MBS shown in Fig. \ref{fig:Conductance_modulated_disorder_SGM_both_side}(h). We therefore emphasize the potential benefit of the SGM technique over the cross-correlated measurements.  

\section{Discussion}
The SGM technique has been applied so far to both two-dimensional \cite{doi:10.1126/science.289.5488.2323, Topinka2001, PhysRevB.80.041303, Iagallo,  Aoki, LeRoy, 10.1063/1.1484548} and quasi-one-dimensional semiconductors such as nanowires \cite{10.1063/1.2746422, Zhukov_2019}. The proximity effect in semiconducting nanowires is typically achieved by introducing a superconducting shell that partially covers the exterior of the nanowire \cite{Mourik, Zhang2018, Heedt2021, https://doi.org/10.1002/adfm.202102388} deposited on the substrate (which is dielectrically separated from the underneath gates). The superconducting metallic cover that envelopes the wire could in principle prohibit SGM probing. The technique that we propose in this manuscript could potentially be used for partially covered nanowires recently developed using shadow wall techniques. These structures have a few planes of the wire covered by a metallic superconductor while leaving a few facets with an exposed semiconductor. They are gateable \cite{https://doi.org/10.1002/adfm.202102388, Heedt2021} as the open part of the wire allows for local potential tuning. Importantly, it has been demonstrated that the tip does not have to be exactly above the wire and even if it is located on its side, the electrostatic potential of the charged tip locally affects the potential distribution in the wire \cite{Bleszynski2007}.

Another system with potential applications for SGM probing, is two-dimensional structures, where a significant surface of the normal region remains uncovered. This perspective is in line with recent progress in the realization of proximitized heterostructures that host two-dimensional electron gases, realized using 2DEGs confined in semiconducting flakes \cite{chieppa2025superconductingquantuminterferencedevices, https://doi.org/10.1002/pssb.202400534, Turini2022, PhysRevResearch.5.033015, Salimian}, or heterostructures
\cite{PhysRevLett.124.226801, PhysRevB.107.245304, Moehle2022, PhysRevLett.130.116203, Fornieri}. Those systems consist of an uncovered semiconductor that can be subjected to the SGM probing and a nearby superconductor that provides electron-hole pairing. These structures provide a straightforward way of application of SGM in a similar fashion to previously studied nanoscopic semiconducting devices. In fact, SGM measurement of a narrow two-dimensional semiconducting Josephson junction has been recently  successfully performed \cite{lombardi2025supercurrentmodulationinsbnanoflagbased}. 

Furthermore, the two-dimensional systems allow for the proximitization of the normal region by more than a single superconductor, allowing for the creation of Josephson junctions with non-trivial geometries \cite{Fornieri, Moehle2022, PhysRevLett.130.116203, PhysRevB.107.245304}. Especially, those systems are promising in the context of probing the appearance of MBSs as they allow the reduction of the necessary magnetic field for the topological transition by control over the superconducting phase difference \cite{PhysRevLett.118.107701, PhysRevX.7.021032} or the dimensions of the junction \cite{PhysRevLett.125.086802, Kuiri}.

\section{Summary and conclusions}
We studied the scanning gate microscopy technique as a tool to detect Majorana bound states in a quasi-one-dimensional semiconducting wire with induced superconductivity. We showed that the scanning gate microscopy applied to such systems can induce the localization of an additional Majorana pair and that the change of position of the probe changes the overlap between Majorana bound states located on each side of the tip. This in turn induces oscillation in the energy structure of the system that can be detected by usual conductance spectroscopy measurements. Moreover, we showed that the scanning gate microscopy technique allows one to detect not only the periodic oscillation of energy of Majorana bound states but also their second fundamental characteristic---the exponential increase of the energy splitting when the span of the system available for Majorana states becomes limited. We also showed that scanning gate probing reveals the difference between the quasi- and true-Majorana bound states resulting from the strong localization of the former at a smooth potential barrier, causing the lack of the oscillations of their energy levels as the system is probed by the tip. We demonstrated that for weaker tip potentials, the two neighboring Majorana bound states hybridize, which opens pronounced anticrossings in the energy spectra, which are reflected in the local conductance maps and which result in non-zero non-local conductance features. We showed that in disordered systems with a mean free path smaller than the length of the system, the oscillations are most pronounced when the length between the tunneling probe and the tip is comparable to the mean free path. Finally, for strongly disordered wires, where the determination of the nature of zero bias peaks can be ambiguous, the scanning gate microscopy technique can be used to distinguish their topological or trivial origin.

\section{Acknowledgments}
This work was supported by the National Science Center, Poland (NCN) agreement number UMO-2020/38/E/ST3/00418. We gratefully acknowledge Polish high-performance computing infrastructure PLGrid (HPC Center: ACK Cyfronet AGH) for providing computer facilities and support within computational grant no. PLG/2024/017374.

\bibliography{reference}
\end{document}